\documentclass[aps,prl,reprint,superscriptaddress,showpacs]{revtex4-1}

\usepackage{color}

\usepackage{graphicx}
\usepackage{booktabs}
\usepackage{bm}
\usepackage{longtable}
\usepackage{amsmath}
\usepackage{dcolumn}
\usepackage{placeins}
\usepackage{expdlist}

\usepackage{siunitx}

\begin{document}

\title {Dual frequency caesium spin maser}

\author{P. Bevington} 
\affiliation{National Physical Laboratory, Hampton Road, Teddington TW11 0LW, United Kingdom}
\affiliation{Department of Physics, University of Strathclyde, Glasgow G4 0NG, United Kingdom}
\author{R.\, Gartman}
\affiliation{National Physical Laboratory, Hampton Road, Teddington TW11 0LW, United Kingdom}
\author{Y. V. \, Stadnik}
\affiliation{Kavli Institute for the Physics and Mathematics of the Universe (WPI), The University of Tokyo Institutes for Advanced Study, The University of Tokyo, Kashiwa, Chiba 277-8583, Japan}
\author{W.\, Chalupczak}
\affiliation{National Physical Laboratory, Hampton Road, Teddington TW11 0LW, United Kingdom}

\date{\today}

\begin{abstract}
Co-magnetometers have been validated as valuable components of the atomic physics toolbox in fundamental and applied physics. So far, the explorations have been focused on systems involving nuclear spins. Presented here is a demonstration of an active alkali-metal (electronic) system, i.e. a dual frequency spin maser operating with the collective caesium F=3 and F=4 spins. The experiments have been conducted in both magnetically shielded and unshielded environments. In addition to the discussion of the system's positive feedback mechanism, the implementation of the dual frequency spin maser for industrial non-destructive testing is shown. The stability of the F=3 and F=4 spin precession frequency ratio measurement is limited at the $3 \times 10^{-8}$ level, by the laser frequency drift, corresponding to a frequency stability of $\SI{1.2}{\milli\hertz}$ for $10^4$ sec integration time. We discuss measurement strategies that could improve this stability to nHz, enabling measurements with sensitivities to the axion-nucleon and axion-electron interactions at the levels of $f_a/C_N \sim 10^9$ GeV and $f_a/C_e \sim 10^8$ GeV, respectively. 
\end{abstract}

\maketitle

The idea of an alkali-metal spin maser system was introduced in the sixties \cite{Bloom1962, Bell1966}. When exploring self-oscillating radio-frequency (rf) atomic systems, Arnold L. Bloom pointed out that ``the response of the spin system with regard to the feedback loop is similar to that of a spin system in a maser (in which case the feedback loop consists of the radiation reaction inside the cavity)''. 
As in other maser (or laser) systems, the alkali-metal spin maser includes population inversion in the form of a spin polarization along an offset magnetic field (i.e. population imbalance) within the ground state manifold. Spontaneous fluctuations create spin components orthogonal to the offset field (i.e. atomic coherences) that precess around the magnetic field at the Larmor frequency. The collective atomic spin precession is optically monitored with a photodetector, the output of which is fed into rf coils located in the vicinity of the atomic vapour. The rf field generated creates a positive feedback signal to the spontaneously generated spin precession provided that the sum of all phase shifts in the feedback loop is zero. Initial studies of this system were limited to the area of geomagnetic measurements \cite{Dyal1969, Kubo1972, Kubo1973, Yabuzaki1974}.
Extensive studies of spin maser properties have been performed in noble gases, where the term spin (or Zeeman) maser was coined and the idea of dual frequency spin maser (DFSM) was developed \cite{Slocum1971, Richards1998, Chupp1994, Bear1998, Stoner1996}.

\begin{figure}[h!]
\includegraphics[width=\columnwidth]{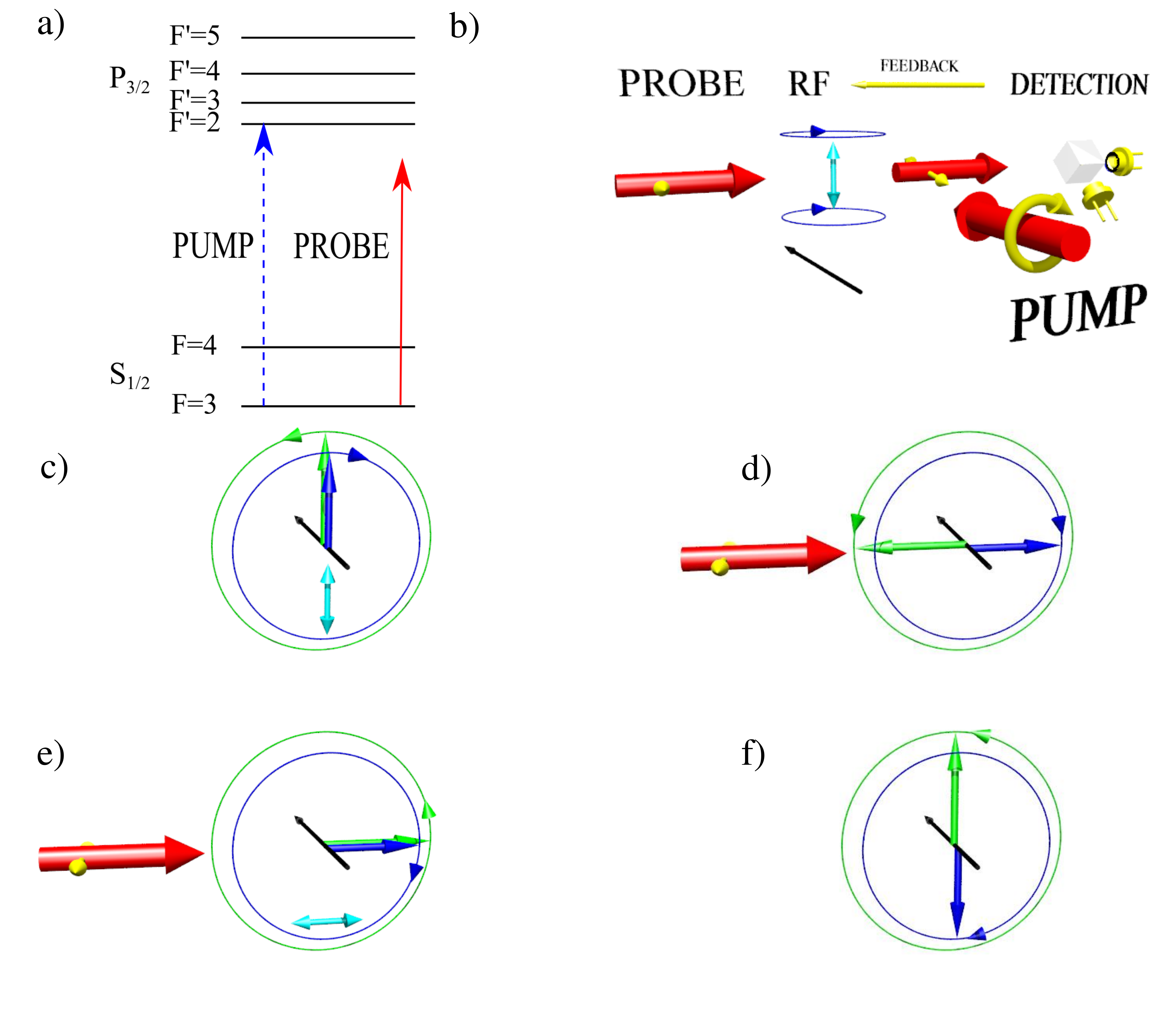}
\caption{(colour online) (a) Caesium D2 line (852 nm) energy structure (pump and probe laser ferquencies marked with dashed blue and solid red lines). (b) Spin maser operating simultaneously at the F=3 and F=4 Larmor frequencies. The F=3 and F=4 atomic spins are oriented by the pump beam along the offset magnetic field (black arrow). (c)-(f) Evolution of the atomic spins in two configurations of the rf field. Green and blue arrows represent components of the F=3 and F=4 spins (coherences) precessing in the offset magnetic field. The rf field (cyan arrow) creates these components in the same direction (c), (e). The two components then proceed to precess in opposite directions  (d), (f). When there is a non-zero projection of the spins on the probe beam axis, a non-zero signal is observed.}\label{fig:Spin_Maser}
\end{figure}

The concept of a DFSM, i.e. an active system with two spins simultaneously precessing in the same offset field, enables relative measurements and addresses the issue of drifts in the operating (Larmor) frequency. Consequently, this expands the area of possible applications beyond basic magnetic field measurements. DFSM and co-magnetometers (in this context, passive systems without a feedback loop) are particularly attractive for inertial sensing \cite{Kitching2011} and testing of fundamental physics \cite{Abel17, Sato2018, Wu2019}.  In noble gasses, DFSM has been achieved in mixtures of $^3$He and $^{129}$Xe \cite{Chupp1994, Bear1998, Stoner1996} or $^{129}$Xe and $^{131}$Xe \cite{Sato2018}. While operation at two frequencies makes the active and passive systems insensitive to drifts in the offset magnetic field, there remains a whole spectrum of fundamental and technical shifts that compromise the frequency ratio measurement. From this perspective, a particularly interesting demonstration was the operation of a co-magnetometer within the same molecule \cite{Wu2018}. The operation of a co-magnetometer or spin maser within identical molecules or atoms, rather than overlapping ensembles of different atoms, reduces the noise contribution due to magnetic field gradients.

We explore a spin maser simultaneously operating with the F=3 and F=4 ground state hyperfine levels components of the collective spin of an ensemble of caesium atoms. Adjustment of the beam power (pump and probe) and the probe detuning from the relevant transitions enables the creation of a signal with equal amplitudes and phases for the F=3 and F=4 spins, despite them precessing in opposite directions. In the following, we discuss in detail the mechanism of the positive feedback, the basic properties of the maser and explore possible applications. The experiments were performed in both magnetically shielded and unshielded environments, demonstrating the system capabilities in both fundamental studies and industrial non-destructive testing (NDT), respectively.
The latter is based on magnetic inductive measurements with an electrically conductive or magnetically permeable object.  The object's response to an oscillating rf magnetic, so-called primary, field indicates the presence of structural defects and composition inhomogeneities within the studied sample. Implementation of the spin maser concept helps to overcome some of the important technological challenges in NDT with rf atomic magnetometers, in particular the stability of the offset field in the presence of a ferromagnetic object and the image acquisition time \cite{Bevington2018, Bevington2019, Bevington2019b}.

The measurements described here are performed in two configurations, shielded \cite{Chalupczak2012a, Chalupczak2012} and unshielded \cite{Bevington2018, Bevington2019, Bevington2019b, Bevington2019c}, both of which have been described before and here only their essential elements are recalled. In both setups, caesium atomic vapour is housed in a paraffin-coated cell at ambient temperature (atomic density $n_{\text{Cs}}=0.33 \times10^{11} \text{cm}^{-3}$). Pumping is performed by a circularly polarised laser beam, frequency stabilized to the $6\,^2$S$_{1/2}$ F=3$\rightarrow{}6\,^2$P$_{3/2}$ F'=2 transition (D2 line, $\SI{852}{\nano\meter}$), Fig.~\ref{fig:Spin_Maser} (a), propagating along the direction of the offset static magnetic field, (b). In the shielded setup, the offset field is produced by a solenoid. A pair of Helmholtz coils around the vapour cell creates the driving rf field. In the unshielded setup, the offset field is defined by a set of nested, orthogonal, square Helmholtz coils. The rf field is generated by a single coil. The probe laser frequency is $\SI{2.75}{\giga\hertz}$ red detuned from the $6\,^2$S$_{1/2}$ F=3$\rightarrow{}6\,^2$P$_{3/2}$ F'=2 transition. The resulting signal is either measured by a lock-in amplifier or recorded by a 2 MS/s data acquisition board.

\begin{figure}[h!]
\includegraphics[width=\columnwidth]{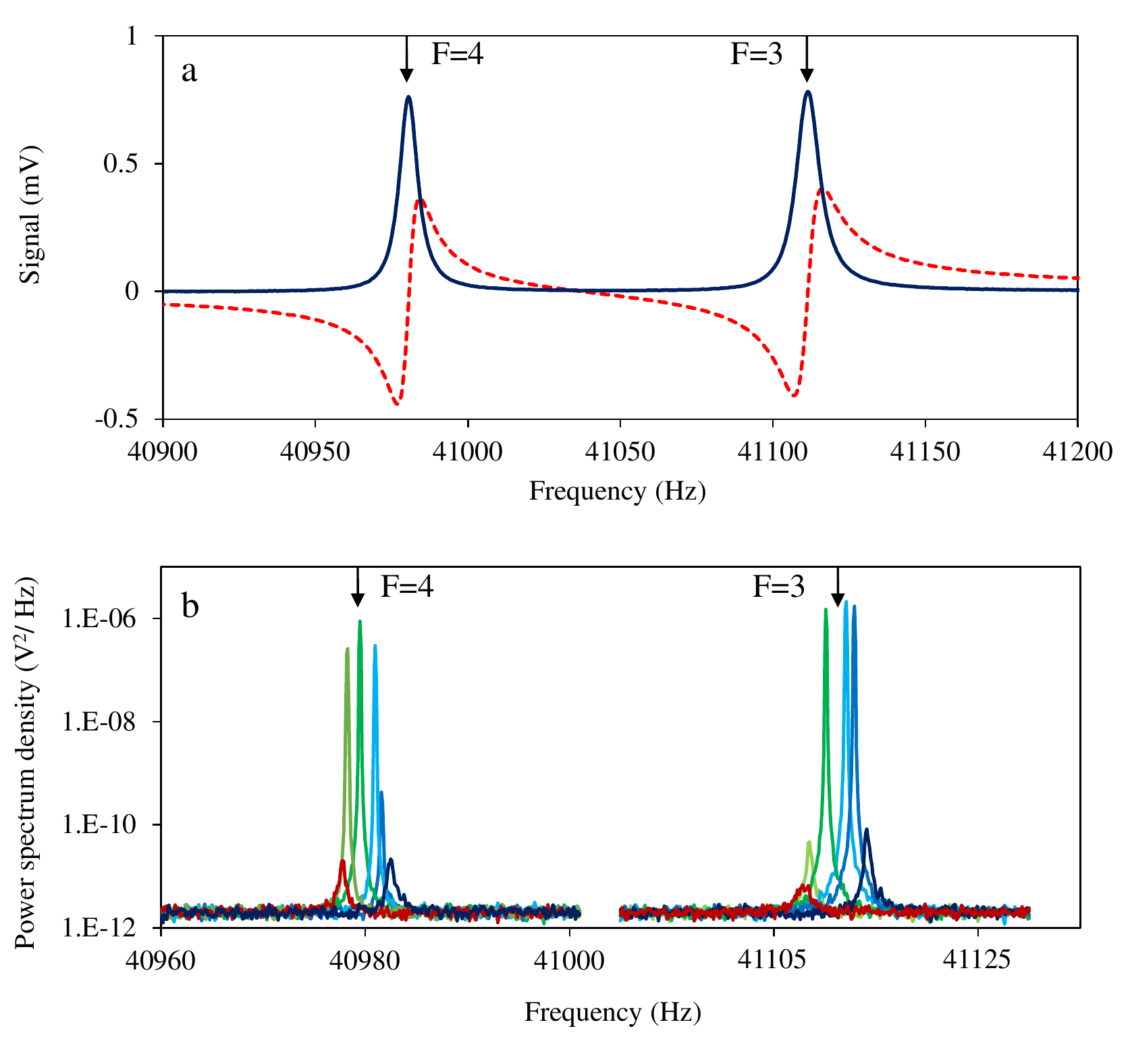}
\caption{(colour online) (a) The spectrum of the rf transition within the F=4 and F=3 ground state levels (marked with black arrows) for an offset magnetic field of 4.5 $\mu$T.  Solid (blue) and dashed (red) lines represent the in-phase and quadrature components recorded by the lock-in amplifier. (b) Power spectral density of the DFSM signal showing a change of the operating frequency for six values of the phase shift in the feedback loop. Red and dark blue line profiles represent amplified spontaneous fluctuations. }\label{fig:Spectrum}
\end{figure}

Radio frequency spectroscopy is used to monitor the spin polarization of the F=3 and F=4 ground states \cite{Chalupczak2012a}. Figure~\ref{fig:Spectrum} (a) shows a typical polarization-rotation signal measured as a function of the rf field frequency. Resonances are observed when the rf field frequency matches the splitting between neighbouring Zeeman sublevels introduced by the offset field. The difference in Land\'{e} factors for the F=3 and F=4 hyperfine ground-states results in opposite signs of Larmor frequency and a separation in respective resonances frequencies. Most of the phase change ($\approx \pm 90^{\circ}$) occurs within the frequency span between the minimum and maximum of dispersive-like component. 
The pump beam is directly coupled to the F=3 level, making the precession of the F=3 spin component prone to perturbations caused by instabilities in the laser amplitude and frequency. Implementation of low power pumping and laser power stabilisation partially addresses this problem. The pumping of the F=4 collective spin is achieved by off-resonant excitation and spin-exchange collisions, resulting in a decoupling of the F=4 spin precession frequency from the laser light.

The operation of a DFSM requires the presence of positive feedback for both the F=3 and F=4 spins. For the case where the rf field axis is orthogonal to the probe beam propagation direction, the photodetector signal must be shifted by $90^{\circ}$ to provide feedback. If they are parallel, no phase shift is necessary.
The detected signal is a product of the amplitude of the atomic spin component along the probe beam direction and the sign of the detuning from the relevant atomic resonance. The positive feedback condition for both spins amounts to selecting the right sign of probe beam detuning from the F=3 and F=4 resonances for a specific direction of the rf field (cyan arrow in Fig.~\ref{fig:Spin_Maser}). We first consider the case where the rf field direction is orthogonal to the probe beam. The F=3 and F=4 atomic spins are oriented by the pump beam along the offset magnetic field, and subsequently tilted by the rf field in the same direction, Fig.~\ref{fig:Spin_Maser} (c). The two components then proceed to precess in opposite directions, hence the projection of either spin component on the probe beam axis have opposite signs, Fig.~\ref{fig:Spin_Maser} (d). If the probe beam frequency is set between the F=3 and F=4 resonances, their detunings signs are opposite. Hence, the output signals for the F=3 and F=4 components have the same sign. For the case where the rf field direction is parallel to the probe beam, the DFSM operation requires the same sign of detuning for both the F=3 and F=4 atoms, Fig.~\ref{fig:Spin_Maser} (e)-(f).
The operation of a DFSM for various rf field geometries and probe beam detunings will be illustrated through the analysis of the spin maser action in an NDT measurement.

\begin{figure}[tbp]
\includegraphics[width=\columnwidth]{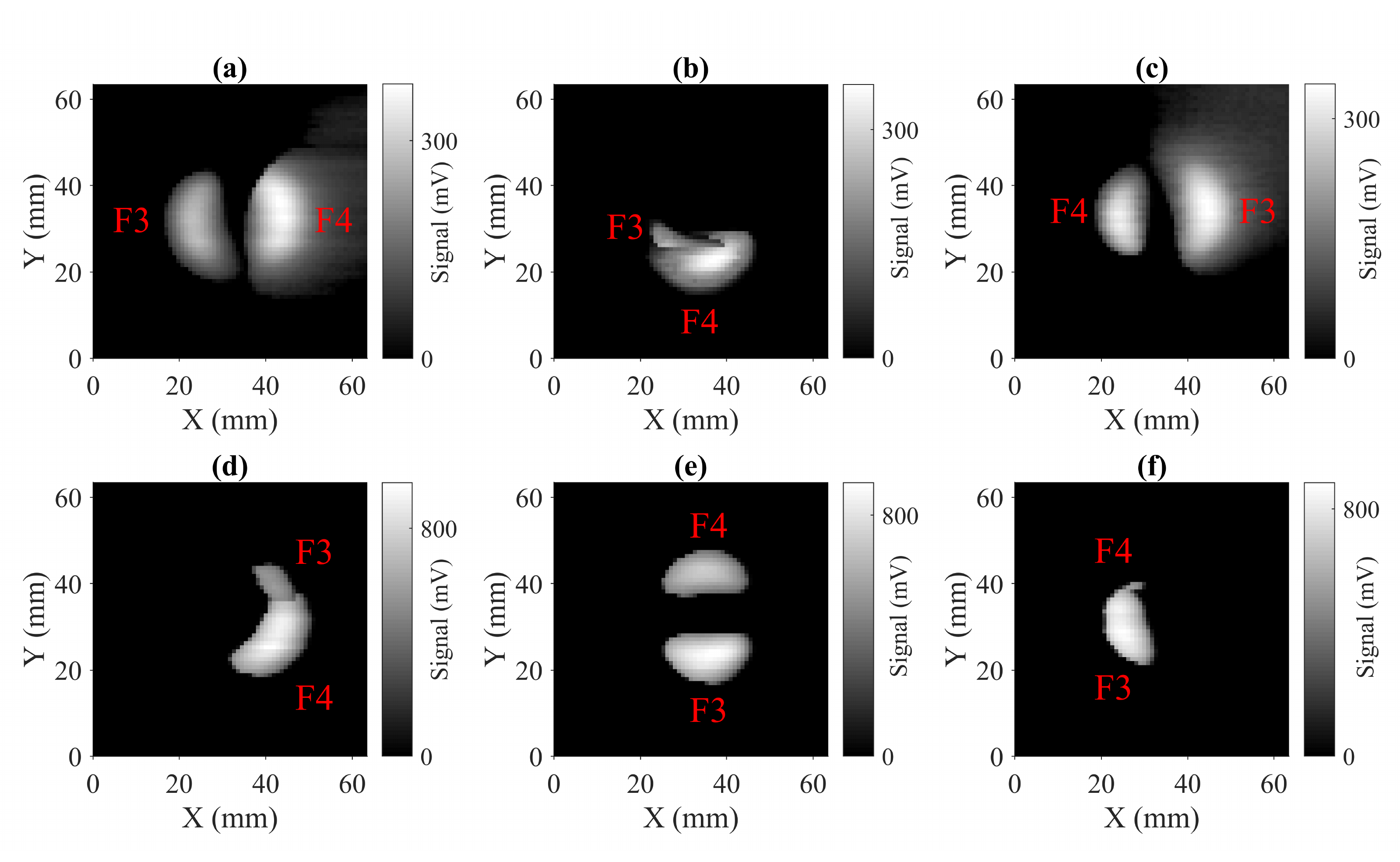}
\caption{Change in dual frequency maser signal amplitude recorded over a $\SI{64}\times\SI{64}{\milli\meter\squared}$ area of a \SI{6}{\milli\meter} thick aluminium plate containing a $\SI{24}{\milli\meter}$ diameter recess that is $\SI{2.4}{\milli\meter}$ deep for $0^{\circ}$  (a), (d) $90^{\circ}$ (b), (e) and $180^{\circ}$ (c), (f) phase shifts in the feedback loop. Images (a)-(c) were recorded with the probe $\SI{2.75}{\giga\hertz}$ red detuned from $6\,^2$S$_{1/2}$ F=3$\rightarrow{}6\,^2$P$_{3/2}$ F'=2 transition. Images (d)-(f) were recorded with the probe beam frequency $\SI{890}{\mega\hertz}$ blue detuned from the $6\,^2$S$_{1/2}$ F=3$\rightarrow{}6\,^2$P$_{3/2}$ F'=2 transition. Measurements were performed with a Larmor frequency $\sim \SI{40}{\kilo\hertz}$.}\label{fig:Phase2}
\end{figure}

The feedback in inductive NDT experiments is created by the response of the tested object (secondary field) to the rf driving (primary) field \cite{Bevington2019c}. Because of the measurement geometry, i.e. the direction of the offset field parallel to the primary field axis and orthogonal to the surface of the tested object, the atomic spin precession can only be driven by the secondary field components parallel to the object surface. As shown in \cite{Bevington2019b}, these field components have non-zero amplitudes only in the vicinity of a defect (recess), where the angular distribution of the secondary field spans the full $360^{\circ}$ range. This angular change maps linearly to the phase of the feedback loop ensuring that the phase-matching condition will always be met over some section of the defect. However, this is a downside of spin maser opperation when compared to the free-running mode (i.e. with an external drive for the rf field) \cite{Bevington2019b}, since it does not record the full defect signature \cite{Bevington2019c}. Application of the DFSM helps to recover a larger section of the defect signature.

The images in Fig.~\ref{fig:Phase2} show the change of the DFSM amplitude recorded over an aluminium plate containing a circular recess.   
The columns in the image array represent three different phase shifts in the feedback loop ($~0^{\circ}$, $~90^{\circ}$, and $~180^{\circ}$) with the two rows reflecting the two different probe beam detunings ($-\SI{2.75}{\giga\hertz}$ and $\SI{890}{\mega\hertz}$) from the $6\,^2$S$_{1/2}$ F=3$\rightarrow{}6\,^2$P$_{3/2}$ F'=2 transition. The different phase shifts applied in the feedback loop are equivalent to the two configurations shown in Fig.~\ref{fig:Spin_Maser}. In particular, a $0^{\circ}$ phase shifts represent the field geometry in  Fig.~\ref{fig:Spin_Maser} (e), while a $90^{\circ}$ phase shifts are equivalent to the field geometry in  Fig.~\ref{fig:Spin_Maser} (d).
Change of the phase shift and the probe laser detuning result in the F=3 and F=4 spin maser operation being triggered at different sections of the recess edge.
 
The opposite direction of the F=3 and F=4 spin precession is mirrored by the opposite phase dependence of the position where spin maser action is achieved.
The phase shift value in Fig.~\ref{fig:Phase2} (a), (d) corresponds to $0^{\circ}$, i.e. the drive acting on the atoms produced by the secondary field is directed along the probe beam, Fig.~\ref{fig:Spin_Maser} (e). For Fig.~\ref{fig:Phase2} (a), because of the opposite signs of the probe beam detuning for F=3 and F=4 atoms, the projections of the F=3 and F=4 spins onto the probe beam axis produce a signal with opposite signs. Consequently, the phase-matching condition for the F=3 and F=4 spins are met at opposite sides of the recess. 
For Fig.~\ref{fig:Phase2} (d), the probe beam frequency results in the same sign of detuning from the F=3 and F=4 resonances and the phase-matching condition is met over the same section of the recess edge for both spins. Figure~\ref{fig:Phase2} (b), (e) represents the case with a $90^{\circ}$ phase shift, i.e. the drive acting on the atoms produced by the secondary field is orthogonal to the probe beam, Fig.~\ref{fig:Spin_Maser} (c). The F=3 and F=4 spins are tilted by the rf field in the same direction but precess clockwise and counter-clockwise, resulting in opposite projections along the probe beam axis, Fig.~\ref{fig:Spin_Maser} (d). This time for Fig.~\ref{fig:Phase2} (b), the detuning of the probe beam results in the same sign of the signal for both components and the F=3 and F=4 spin maser action regions overlap. For Fig.~\ref{fig:Phase2} (e), the probe beam has the same sign detuning and results in the F=3 and F=4 phase-matching conditions being met at opposite sides of the recess. The increase of the phase shift to $180^{\circ}$, Fig.~\ref{fig:Phase2} (c), (f) results in the reversal of the image shown in Fig.~\ref{fig:Phase2} (a), (d). 
We note that in regions of overlapping F=3 and F=4 maser action, FFT measurements indicate that the system only oscillates at a single frequency and shows behaviour typical for a bistable system. This effect has also been observed in a shielded system where the feedback is produced directly by the rf coils. We ascribe it to the coupling between two oscillating modes \cite{Agrawal1981, Raithel1995, Emelianova2014} and this aspect of the system behaviour will be discussed elsewhere.

Figure~\ref{fig:Spectrum} (b) shows the power spectral density of the maser signal recorded for various values of the phase shift in the feedback loop.
As with any driven oscillator, the phase of the atomic response varies by $180^{\circ}$ if the frequency is tuned across resonance. The spin maser will operate at whichever frequency leads to an overall phase shift of zero \cite{Kubo1973, Yabuzaki1974}.
Consequently, as shown in Fig.~\ref{fig:Spectrum} (b) the operating frequency changes for different external phase shifts with the rate set by linewidth, Fig.~\ref{fig:Spectrum} (a). This is particularly important when considering a frequency ratio measurement, which is addressed in the following paragraph. For large phase shifts ($\approx 90^{\circ}$) the response to the drive field is manifested as a broad, low amplitude peak in the power density spectrum. This is the spin equivalent of amplified spontaneous emission in laser systems. The signal is created by spin fluctuations that are fed into the rf coil with positive feedback that has a strength comparable with the decoherence rate. The frequency spread of the signal is defined by the decoherence rate and offset field inhomogeneity.

\begin{figure}[h!]
\includegraphics[width=\columnwidth]{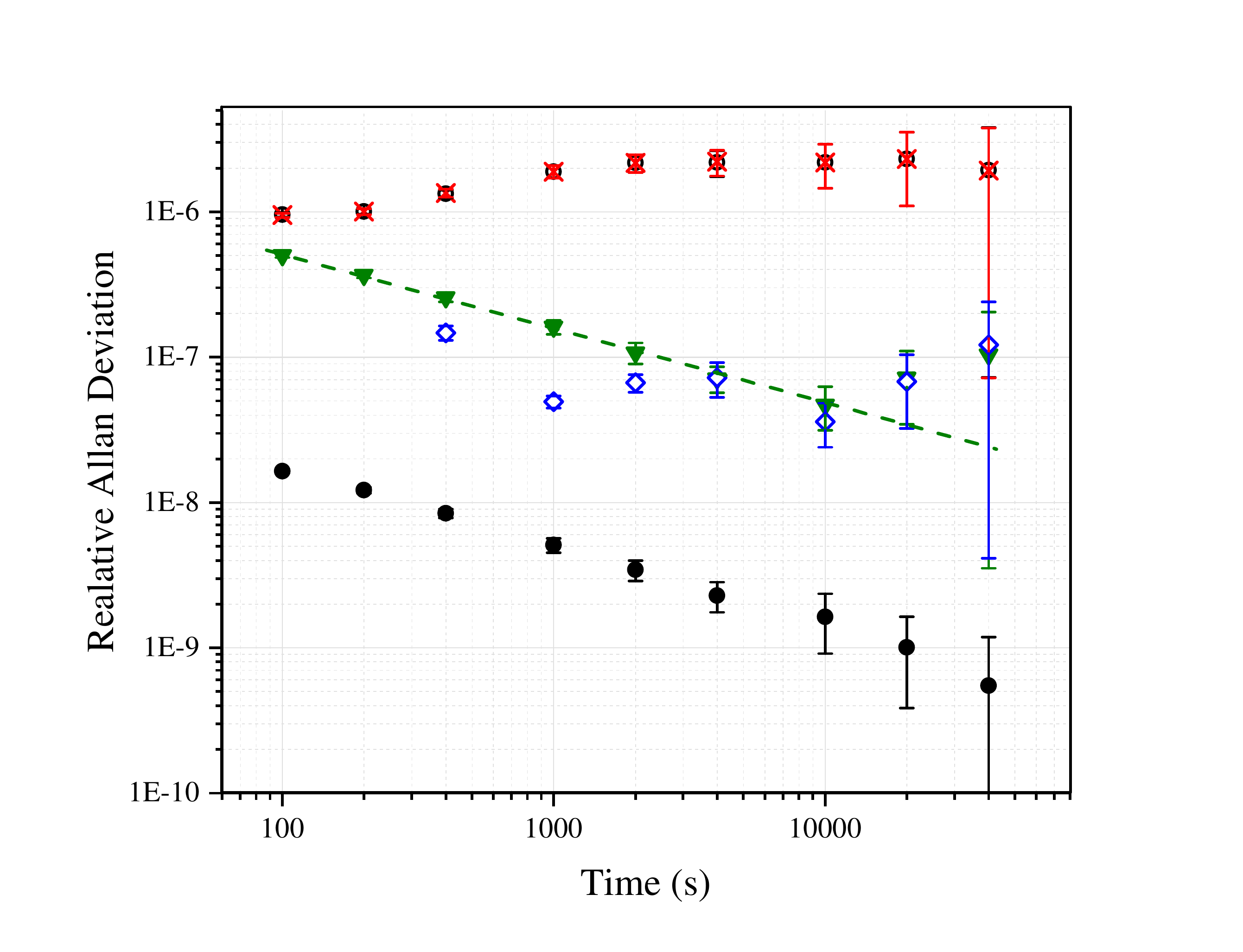}
\caption{The statistical uncertainty (Allan deviation) of the frequency of the F=3 and F=4 spin maser components (black circles and red crosses) recorded in the shielded setup. The frequencies are normalised to the initial value. The behaviour of the Allan deviation indicates flicker noise, which result from the temperature driven variations in the magnetic field. The initial dependence of the statistical uncertainty of the frequency ratio (green triangles) on integration time ( $\sim \tau^{-\frac{1}{2}}$) indicates a dominant `white noise' character. The black points and blue diamonds represent the limits imposed by the two leading contributions, the pump beam amplitude and frequency stability.
The offset magnetic field is set to $11.4 \mu$T, Larmor frequency $\approx \SI{40}{\kilo\hertz}$.}\label{fig:Allan}
\end{figure}

Implementation of co-magnetometers in anomalous spin-dependent interactions studies (e.g. coupling with dark matter) involves searching for a time-dependent signal in the measurement of the Larmor frequency ratio of two species \cite{Stadnik2017, Abel17, Wu2019, Bloch2019},  as well as time-independent signals \cite{Vasilakis2009, Fadeev2019}. 
We discuss this measurement, i.e. the measurement of the ratio between the F=3 and F=4 spin precession frequencies, in order to illustrate the capabilities and challenges of implementing a DFSM in fundamental physics explorations. The measurement is divided into $100 \, \text{sec}$ long sections. 
The resonance frequencies of the F=3 and F=4 components are extracted from the FFT spectrum of each section. 
Figure~\ref{fig:Allan} shows the statistical uncertainty (Allan deviation) of the frequency of the F=3 and F=4 spin maser components (black circles and red crosses) as well as their ratio (green triangles). In order to show all the results in one plot, the frequencies of the F=3 and F=4 components have been normalised to their initial value. The short term stability of the F=3 and F=4 normalised frequency measurement is set at $\approx 2 \times 10^{-6}$ level by the stability of the current source. For  $\tau < 10^{3} \, \text{sec}$ the behaviour of their statistical uncertainty indicates flicker noise in the measurement data and the long-term behaviour reflects drifts in the magnetic field caused by temperature variations.
The frequency ratio (green triangles) is not sensitive to variations in the magnetic field and it follows a $\sim \tau^{-\frac{1}{2}}$ trend (dashed line) for $\tau < 10^4 \, \text{sec}$, indicating a dominant `white noise' character. 
Evaluation of the stability limits involves: (1) a measurement of the slope of the frequency ratio dependence on various measurement parameters, such as laser beam power and laser frequency stability, and (2) monitoring the parameters' value over the course of the frequency ratio measurements. While the long term stability of the F=3 to F=4 frequency ratio is not affected by the laser beams powers (dependence on the pump beam power represented by black dots), the Allan deviation of the frequency ratio (green triangles) for $\tau > 10^4 \, \text{sec}$ mirrors that of the limit set by the pump laser frequency noise (blue diamonds). The laser frequency drift limits the stability of the F=3 and F=4 spin precession frequency ratio measurement at the $3 \times 10^{-8}$ level, that corresponds to a frequency stability of $\SI{1.2}{\milli\hertz}$ for $10^4 \, \text{sec}$ integration time. 
The drifts in the laser frequency stabilisation is an artefact of the non-optimised temperature stabilisation unit in the laser diode controller \cite{Footnote1}. 
We verified that fitting a waveform to the raw signal (with the signal-to-noise level, SNR $\sim 10^2$) can improve the accuracy of the frequency measurement by two orders of magnitude, compared with the limit defined by the FFT resolution. Exploiting this allows us to reach an uncertainty in the frequency ratio of $\approx 10^{-10}$, equivalent to the frequency stability at $\approx \SI{4}{\micro\hertz}$ level after $10^4 \, \text{sec}$ integration time \cite{Sato2018, Wu2019}. In the context of Dark Matter searches \cite{Stadnik2014}, this would translate into sensitivities to the axion-nucleon and axion-electron interactions at the levels of $f_a/C_N \sim 10^5$ GeV and $f_a/C_e \sim 10^4$ GeV, respectively, for the axion masses $m_a \lesssim 10^{-18}$ eV/$c^2$. 

In conclusion, we have demonstrated the operation of a caesium DFSM in magnetically shielded and unshielded environments. We have discussed the various rf/ laser field configurations necessary for the observation of the maser action. While further improvements of the setup are possible, the direct optical coupling to the F=3 state makes the presented system not ideal for a frequency ratio measurement. Use of the two naturally occurring isotopes of Rb, with indirect pumping configured as in Ref. \cite{Chalupczak2012a}, would increase the SNR ($\sim 10^4$) as indirect optical pumping would enable stronger atomic spin polarisation. This would also reduce the sensitivity on the pump beam frequency and power stabilities. Consequently, the discussed approach would enable measurements with a frequency stability at the nHz level, and therefore, at the levels of $f_a/C_N \sim 10^9$ GeV and $f_a/C_e \sim 10^8$ GeV. This would improve the sensitivity to the axion-electron coupling strength with respect to spin-polarised torsion pendulum experiments \cite{Terrano2019}.

The work was funded by UK Department for Business, Innovation and Skills as part of the National Measurement System Program.  P.B. was supported by the Engineering and Physical Sciences Research Council (EPSRC) (No. EP/P51066X/1). The work of Y.V.S.~was supported by the World Premier International Research Center Initiative (WPI), MEXT, Japan.


\begin{thebibliography}{10}
\bibitem{Footnote1} Temperature oscillation and drift were observed in early models of Vescent laser and require replacement of the temperature stabilisation unit.
\end{thebibliography}
\end{document}